\documentstyle[epsfig]{aipproc}

\def\Msun{~M_\odot}
\def\lsim{\raise0.3ex\hbox{$<$}\kern-0.75em{\lower0.65ex\hbox{$\sim$}}}
\def\gsim{\raise0.3ex\hbox{$>$}\kern-0.75em{\lower0.65ex\hbox{$\sim$}}}

\def\kms{\rm ~km~s^{-1}}

\def\apj{{\it ApJ}}

\begin{document}
\title{Supernova Remnants in Molecular Clouds}

\author{Roger A. Chevalier}
\address{Department of Astronomy, University of Virginia \\
P.O. Box 3818, Charlottesville, VA 22903}

\maketitle

\begin{abstract}
Supernovae are expected to occur near the molecular material
in which the massive progenitor star was born, except in cases where
the photoionizing radiation and winds from the progenitor
star and its neighbors have cleared out a region.
The clumpy structure in molecular clouds is crucial for the
remnant evolution; the supernova shock front can become radiative
in the interclump medium and the radiative shell then collides
with molecular clumps.
The interaction is relevant to a number of phenomena:
the hydrodynamics of a magnetically supported dense shell interacting with
molecular clumps; 
the molecular emission from shock waves, including
the production of the OH 1720 MHz maser line; the relativistic particle
emission, including radio synchrotron  and $\gamma$-ray emission,
 from the dense radiative shell; and the possible
gravitational instability of a compressed clump.

\end{abstract}

\section*{Introduction}

Massive stars are born in molecular clouds and those with masses
$\gsim 8\Msun$ end their lives as supernovae after $\lsim 3\times 10^7$ years.
This age is less than the typical age of molecular clouds so that the
supernovae may explode in the vicinity of molecular cloud material.
Early studies of the interaction assumed that the supernovae
interacted directly with molecular cloud material, with density
$n_H=10^4 - 10^5$ cm$^{-3}$ \cite{S80},\cite{WMS80}.
At these high densities, the supernova remnant evolves on a timescale
of 10's of years during which it is a luminous infrared source.

Such sources have not been clearly observed.
However, supernova remnant interaction with molecular gas has
been observed, starting with the remnant IC 443 \cite{deN79b}.
Continued observations have shown IC 443 to be one of the best cases
of molecular cloud line emission \cite{vD93}.
The emission appears to be associated with shock interaction with
dense clumps, as expected in a molecular cloud;
   the point of view adopted here involves the expansion of
the supernova remnant shell in the interclump medium of a cloud
with some collisions with clumps \cite{C99}.

\section*{The Supernova Surroundings}

A crucial point for the evolution of supernova remnants in molecular
clouds is the clumpy structure of these clouds.
The basic picture that has emerged is of dense clumps with most
of the mass embedded in a lower density interclump medium.
Blitz \cite{B93} notes some properties of typical 
Giant Molecular Clouds:  total mass $\sim 10^5\Msun$, diameter
$\sim 45$ pc,  H$_2$ density in clumps of $10^3$ cm$^{-3}$,
 interclump gas  density in the range $5-25$ H atoms cm$^{-3}$,
and clump filling factor of $2-8$\%.
The clumps have an approximate power law mass spectrum, such that
there are more small clumps, but most of the mass is in the
large clumps.
There have been recent attempts to model the  clump
properties in terms of MHD (magnetohydrodynamic) turbulence \cite{V00,W00}.
In this case, the clumps may be transitory features in the
molecular cloud.

If the dense clumps are confined by the interclump pressure,
a pressure $p/k\approx 10^5$ K cm$^{-3}$ is needed \cite{B93}. 
Even if the clumps are features in a fluctuating velocity
field and are not necessarily confined,
 a comparable pressure is needed to support the clouds
against gravitational collapse.
The thermal pressure in the interclump gas is clearly too small and
magnetic fields are a plausible pressure source.
Studies of the polarization of the light from stars behind molecular
clouds implies that the interclump magnetic field does have
a uniform component \cite{H93}.
Setting $B_o^2/8\pi$ equal to the above pressure yields $B_o=19p_{5}^{1/2}~\mu$G,
where $B_o$ is the uniform field component and $p_5$ is $p/k$ in
units of $10^5$ K cm$^{-3}$.
However, a uniform magnetic field would not provide support along the
magnetic field and would not explain the large line widths observed
in clouds.
An analysis of the polarization and Zeeman effect in the dark cloud
L204 shows consistency with equal contributions to the pressure
from a uniform field and a fluctuating, nonuniform component
\cite{H93}.
The uniform component is then $B_o=13p_{5}^{1/2}~\mu$G.
The magnetic field strength deduced in L204 is consistent with
this value.

The molecular cloud can be influenced by stellar mass loss and
photoionization by the progenitor star before the supernova.
The effect of these processes can be to remove molecular gas from
the vicinity of the central star.
Draine and Woods \cite{DW91} found that stars with initial mass
$\lsim 20\Msun$ in clouds of density $n_H\gsim 10^2$ cm$^{-3}$
would have such a small effect that the surroundings could be
treated as homogeneous in discussing the blast wave evolution.
For a $25\Msun$ star, they found the presupernova effects to be
small provided $n_H\gsim 10^3$ cm$^{-3}$.
In an interclump medium with $n_H\approx  10$ cm$^{-3}$,
Chevalier \cite{C99} found that stars with mass $\lsim 12\Msun$
would clear a region with radius $\lsim 5$ pc around the star
because of the effects of photoionization and winds.
The high pressure of the molecular cloud plays a role in limiting
the effects of the progenitor star.
If the massive star has a significant velocity ($\sim 5\kms$),
the star can move into a region of undisturbed cloud material
and the progenitor star effects are lessened \cite{DW91}.
On the other hand, massive stars tend to form in clusters and
the immediate environment of a star could be affected by  nearby,
more massive stars.
In a sparse cluster, it is still possible for a massive star
 to interact with its natal cloud material.

\section*{The Case of IC 443}

The remnant IC 443  shows many of the features that 
characterize molecular cloud interaction.
The remnant appears to be interacting with a relatively low mass
molecular cloud ($\lsim 10^4\Msun$) that is primarily in front
of the remnant \cite{C77}.
This low mass suggests that there were few or no massive stars
which would be likely to disrupt the cloud with their photoionizing
fluxes and winds.
A lower mass Type II supernova could thus interact directly
with the molecular cloud.
I take the distance to the remnant to be 1.5 kpc \cite{FK80}.

The morphology of the remnant is a shell in the molecular cloud
region, with an apparent ``break-out'' region to the southwest
(see \cite{AA94} for an X-ray image and \cite{C97} for a radio image).
I consider the shell part of the remnant  to be
interacting with the molecular cloud.
Fesen and Kirshner \cite{FK80} found that the spectra of the optical filaments
in this region
imply an electron density for the [S~II] emitting region of 
$\lsim 100$ to 500 cm$^{-3}$ and shock velocities in the range
$65-100\kms$.
If the magnetic field does not limit the  compression at a
temperature $\sim 10^4$ K,
the preshock  density implied by the higher density filaments is
$10-20$ cm$^{-3}$.
The filaments with a lower density may have their compression limited by
the magnetic field.
These observations point to a radiative phase of evolution for the
remnant in the interclump medium of the molecular cloud.
The cooling shock front is expected to build up a shell of HI \cite{C74}.
For an ambient density of $15$ cm$^{-3}$ and a radius of 7.4 pc,
the total swept-up mass is $1000\Msun$ for a half-sphere.
HI observations suggest that there
is $\sim 1000\Msun$ of shocked HI in the northeast shell  part of the remnant,
with velocities up to $\sim 110\kms$ \cite{GH79}.
These results are in quantitative accord with expectations for a $10^{51}$
erg supernova in a medium with $n_o=15$ cm$^{-3}$ \cite{C99,C74}, if account
is taken of the fact that some energy has been released in the blow-out
region.

In a restricted region across the central, eastern part of IC 443,
a rich spectrum of molecular emission has been observed
\cite{vD93,R95b}.
A set of spectroscopically distinct CO clumps has been labelled A--G
\cite{deN79b,H86}, although the H$_2$ emission indicates that the
clumps are associated through a filamentary molecular
structure \cite{B90}.
These clumps, with  sizes $\sim$1 pc, have masses of $3.9-41.6\Msun$ as
deduced from  $^{12}$CO lines \cite{D92}.
The corresponding densities, $n_H\lsim 500$ cm$^{-3}$, are low but
should be regarded as lower limits because the $^{12}$CO emission is
assumed to be optically thin.
Also, there is likely to be density structure within individual clumps.
Analysis of absorption in one place yielded a preshock density of
$n({\rm H_2})\approx 3,000$ cm$^{-3}$ \cite{vD93}.
The properties of these clumps are consistent with those seen in
quiescent molecular clouds.
High spatial resolution is possible in the H$_2$ infrared
lines, which show clumps down to a scale of
1$^{\prime\prime}$ ($2\times 10^{16}$ cm$=0.007$ pc at a distance of 
1.5 kpc) but not smaller \cite{R95b}.
The emission has a knotty appearance
that is unlike the filamentary appearance of the optical emission in IC 443
and the H$_2$ emission in the Cygnus Loop \cite{R95b}.
This type of structure is consistent with the possible fractal structure of
molecular clumps \cite{W00}.

The 2MASS survey has provided a $K_s$ band image of IC 443 that is
likely to be dominated by shocked H$_2$ emission \cite{RJ00}.
The image clearly shows a ridge of emission passing from the south
to the central part of the remnant.
The $J$ and $H$ band images with the 2MASS survey show emission from
the northeast part of the remnant, which is likely to be [Fe II]
emission from the $\sim 100\kms$ shock wave.
Observations of the [O I] 63 $\mu$m line with {\it ISO}
(Infrared Space Observatory) show that it is particularly strong
in the northeast shell region \cite{RJ00}, which  implies that
it is also from the $\sim 100\kms$ shock.
This is consistent with expectations that most of the [O I] 63 $\mu$m line
luminosity should be from the shock front in the interclump
medium \cite{C99}.

\begin{figure}[b!] 
\centerline{\epsfig{file=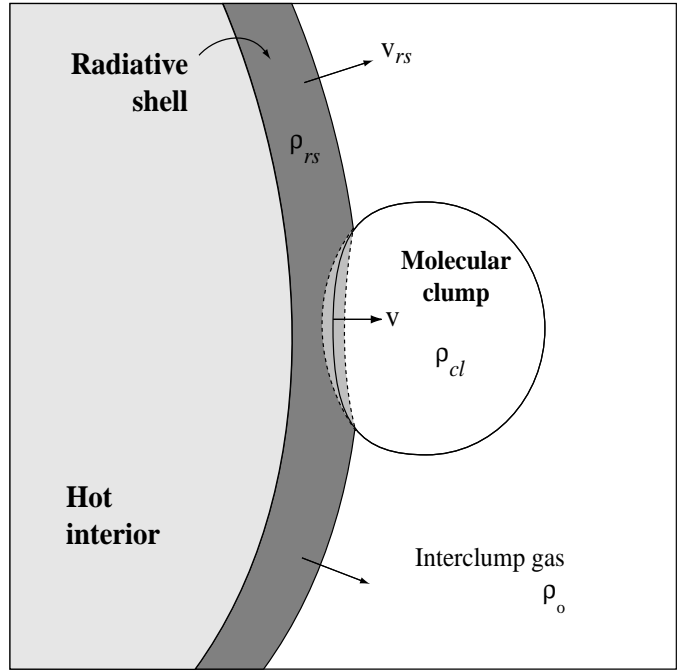,height=3.5in,width=3.5in}}
\vspace{10pt}
\caption{Interaction of a radiative shell,
moving at velocity $v_{rs}$,
with a molecular clump.
The interaction generates a dense slab bounded by shock waves
(dashed lines) and
moving at velocity $v$.
The densities of the molecular clump, $\rho_{cl}$, the radiative
shell, $\rho_{rs}$, and the interclump medium, $\rho_{o}$,
are indicated.   From [5].}
\label{fig1}
\end{figure}

The interpretation of the molecular line strengths is still
controversial, but the emission is most consistent with that from a partially
dissociating $J$(jump)-type shock 
\cite{vD93,R95b,B90,R95a}.
The required shock velocity, $25-30\kms$, is consistent with the
velocities observed in the strongest emission.
The dissociated H$_2$ can be observed as a high column density of
shocked HI at the positions of the clumps \cite{BS86}.
Following \cite{C99}, I interpret the clump shock as being driven by
the radiative shell in the interclump region (Fig. 1);
this can generate a  pressure in the interaction region
significantly above that expected for ram pressure equilibrium.
For a clump shock velocity $v=25-30\kms$ and a radiative shell
velocity $v_{rs}=100\kms$, the ratio of clump density to
shell density $\rho_{cl}/\rho_{rs}$ is
in the range $5.4-9.0$.
The density ratio can vary significantly with relatively little
effect on the shock velocity.
The clump shock velocity is consistent with shell density
$n_{rs}=500$ cm$^{-3}$
and preshock clump density
$n_{cl}=3000$ cm$^{-3}$, which are plausible values.
The ratio of postshock pressure in the clump to that in the
interclump medium (with $n_o=15$ cm$^{-3}$) is 18. 

The column density through the radiative shell is
$N_H\approx n_o R/3\approx 10^{20}$ cm$^{-2}$.
For $n_{cl}=3000$ cm$^{-3}$,
the column density through the clumps is $10^{22} \ell_{\rm pc}$ cm$^{-2}$,
where $\ell_{\rm pc}$ is the path length through the clump in pc.
The larger observed clumps are thus expected to be passed over by the
radiative shell and left in the interior of the remnant.
For clumps with a size $\lsim 0.02$ pc, the shock front breaks out of
the clump first and there is the possibility of further acceleration
by the radiative shell.
The initial shock front through the clump may not dissociate
molecules, which are then accelerated by the shell.
Low column densities of high velocity molecular gas can be produced
in this way.
High velocity molecular gas has been observed in IC 443 at the edge
of one of the clumps \cite{T94}.

Cesarsky et al. \cite{Ce99} have observed H$_2$ rotational lines
lines from clump G with {\it ISO}.
They find that the line fluxes are consistent with a shock velocity $\sim 30
\kms$, a preshock density $\sim 10^4$ cm$^{-3}$, and an evolutionary
time of $1,000-2,000$ years.
At constant velocity, the shock has penetrated $\sim 0.03-0.06$ pc into
the clump.
This result is consistent with the clump interaction scenario.

Another important diagnostic of the molecular interaction is the
OH 1720 MHz maser line, which Claussen et al. \cite{C97} find to
be associated with clump G in IC 443.
Lockett et al. \cite{L99} examined the pumping of the maser and found
that the emission implies temperatures of $50-125$ K and densities
$\sim 10^5$ cm$^{-3}$; the shock must be $C$-type.
An important aspect of the OH maser observations is that it is possible
to estimate the line-of sight magnetic field from the Zeeman effect.
Although they do not have a result for IC 443,
Claussen et al. \cite{C97} find $B_{||} \approx 0.2$ mG for the maser spots in
W28 and W44.
This field strength is compatible with the field strength expected
in the radiative shells of the supernova remnants \cite{C99}.

In addition to the molecular line emission, IC 443 is a source of
nonthermal continuum emission.
At low frequencies, the power law radio spectrum is likely to be 
explained by synchrotron emission.
The observed spectral index, $F_{\nu}\propto \nu^{-0.36}$ \cite{EM85}, implies
a particle energy spectrum of the form $N(E)\propto E^{-1.72}$.
Duin and van der Laan \cite{D75} explained the radio emission by the
shock compression of the ambient magnetic field and relativistic
electrons.
First order Fermi accleration should be included and is consistent with
the observations if ambient cosmic rays are accelerated
\cite{C99,BC82}.
In the shock acceleration of ambient particles, the particle spectrum
is maintained in the postshock region if the particle spectrum is
flatter that $N(E)\propto E^{-2}$.
Chevalier \cite{C99} argued that the relatively flat spectrum may be
consistent with the Galactic cosmic ray electron spectrum, which is observed
to be flat at low energies, perhaps because of Coulomb losses in
the interstellar medium.

If particles are injected into first order shock acceleration at a low
energy, the spectrum tends to $N(E)\propto E^{-2}$ in the postshock
region in the test particle limit.
However, Ostrowski \cite{O99} found that second order Fermi acceleration
in the turbulent medium near the 
shock could give rise to a relatively flat spectrum,
as observed in IC 443.
In another calculation including shock acceleration of electrons from a 
thermal pool, Bykov et al. \cite{B00} found that the combination
of nonlinear effects on the shock compression and free-free absorption of the
radiation could give rise to the observed spectrum (Fig. 2).

\begin{figure}[t!] 
\centerline{\epsfig{file=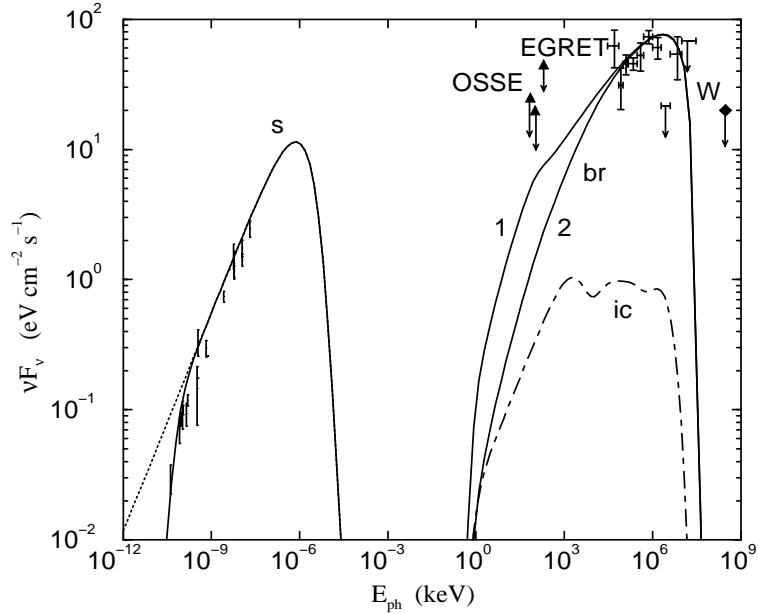,height=3.5in,width=4.2in}}
\vspace{10pt}
\caption{Broadband $\nu F_{\nu}$ spectrum of
the shell of IC 443  calculated from a model of
nonthermal electron production by a radiative shock
with direct injection of electrons from the thermal pool.
The  shock velocity is
150 km s$^{-1}$, the interclump number density is 25 cm$^{-3}$, and
the interclump magnetic field is $ 1.1\times 10^{-5}$ G.
 The two solid curves labeled by numbers 1 and 2
 correspond to two limiting values of an MHD turbulence parameter.
The extended inverse Compton emission  from the whole
remnant is shown 
as a dashed line (ic).
The observational data points
are  from [32] for the {\it EGRET}
source 2EG J0618 +2234 and [27]
for the IC 443 radio spectrum. Upper limits
for MeV  emission are from {\it OSSE} observations
[34]. Upper limits  for $\gamma$-rays $\geq$ 300 GeV 
are from {\it Whipple} observations of IC 443 [36].
From [31].}
\label{fig2}
\end{figure}

An additional constraint on nonthermal particles comes from the 
$\gamma$-ray detection of  IC 443 by the {\it EGRET} experiment on {\it CGRO}
(Compton Gamma-Ray Observatory) \cite{E96},
although the poor
spatial resolution of the {\it CGRO} still leaves some doubt
about the detection.
Assuming that the detection is correct, the emission mechanisms that
might contribute to the radiation are electron bremsstrahlung, pion
decays, and inverse Compton radiation \cite{G98}.
A number of studies have fit the $\gamma$-ray spectrum of IC 443
with these processes \cite{S97,B99}, but they have assumed adiabatic
shock wave dynamics.
In view of the strong evidence for  radiative shocks in IC 443,
Bykov et al. \cite{B00} studied electron injection and acceleration in
the context of  radiative shock waves in a relatively dense medium.
A model with injection of electrons from the thermal pool could
approximately fit the observations, with bremsstrahlung being the
dominant radiation mechanism (see Fig. 2).
As can be seen in Fig. 2, the upper limit on TeV $\gamma$-rays from
the Whipple Observatory \cite{Bu98} places an important constraint
on the high energy part of the particle spectrum.

The continuum radio image of IC 443 \cite{C97} shows that although
there is some correlation with the molecular emission \cite{RJ00}, the shocked
molecular clumps do not stand out as sources of radio synchrotron
emission.
In hard X-ray observations with {\it ASCA}, Keohane et al. \cite{K97}
found two emission regions in IC 443, which have also been found
in {\it BeppoSAX} observations \cite{BB00}.
Recent observations with {\it Chandra} show that the brighter source
is likely to be a pulsar wind nebula \cite{Cl00}, as indicated by earlier
observations \cite{C99}.
The other source may be associated with the molecular interaction
along the southeast rim of the remnant.
Keohane et al. \cite{K97} suggested that the emission could be
synchrotron emission from shock accelerated electrons, but the slow
shock velocity in the radiative shock model makes such an interpretation
unlikely \cite{C99}.
Bykov et al. \cite{B00} suggested that the emission could be bremsstrahlung
from electrons accelerated in ionizing shock waves in relatively
dense gas.
The evidence for $C$-shocks giving rise to OH maser emission implies
 non-ionizing shock waves, but different conditions could prevail
in different clumps.
The prediction is that the shock waves in the interclump gas can produce
bremsstrahlung $\gamma$-ray emission, while the shock waves in the
clump gas can in some cases produce hard X-ray emission.

The radio emission from IC 443 is relatively highly polarized for a
supernova remnant, indicating a preferred direction for the magnetic
field in the shell \cite{K72}.
The field direction agrees with that in the ambient cloud, as determined
by the polarization of starlight.
This is consistent with an ordered magnetic field in the cloud, as
mentioned for molecular cloud support in the previous section.

The ordered field observed in radio emission refers to the field
in the swept up radiative shell.
However, there may also be a uniform component to the magnetic field
in the hot interior of the radiative remnant, which would allow
heat conduction to operate in the interior.
IC 443 is observed to have a relatively isothermal interior with
$T\approx 10^7$ K \cite{P88}.
An isothermal interior with enhanced central X-ray emission is a 
characteristic of a number of remnants that are interacting with
molecular gas.
Shelton et al. \cite{S99} have discussed a conduction model for
the remnant W44, which is a member of this class.
They find that the model X-ray emission distribution profile is close
to that observed, although it is not as centrally concentrated.

\section*{Discussion and Future Prospects}

After a period of   neglect, the study of supernova remnants interacting
with molecular material has entered a phase of more intense study.
In addition to IC 443, a number of remnants have come in for study.
One reason for the increase   is the re-discovery of
the OH 1720 MHz maser line emission as a signpost of molecular
clump interaction.
Although initial observations were made more than 30 years ago \cite{GR68}, 
it was not until after the recent observations by Frail et al. \cite{FGS94}
that the importance of the emission for remnant studies was realized.
Surveys of remnants \cite{Fr96,Gr97} have yielded 17 remnants with the OH maser
line emission out of 160 observed.
In addition, 
Koralesky et al. \cite{Ko98} made VLA observations searching for maser 
emission toward
20 remnants and found shock excited emission in 3 of them.
When remnants with OH maser emission 
are observed in the CO line, molecular clumps
have been observed in a number of cases.
Frail and Mitchell \cite{FM98} found CO clumps associated with the OH emission
in W28, W44, 3C 391 and Reynoso and Mangum \cite{RM00} found such clumps in 
three other
remnants.  

Another source of progress has been the availability of infrared
observations, especially recently with {\it ISO}.
{\it ISO} has been instrumental in detecting atomic fine structure
lines, molecular lines, and dust continuum emission from remnants that
appear to be interacting with molecular gas \cite{RR96,RR98,RR00}.
Of particular interest is the study by Reach and Rho \cite{RR00} of
emission from W28, W44, and 3C 391.
They found evidence for interaction with a clumpy medium in which higher
pressures were attained at higher densities, as discussed for
IC 443.
This situation can be explained by a dense radiative shell interacting
with molecular clumps.
More recently, the 2MASS survey has provided near-infrared imaging
of the sky.
The 2MASS $K_s$ band image of IC 443 is likely to be dominated by shocked
H$_2$ emission \cite{RJ00}.
The detection of supernova remnants in this band thus provides  a good
indication that molecular interaction is taking place.
The advent of the future NASA infrared observatories {\it SOFIA} and
{\it SIRTF} will allow more complete studies of the density structure
in shocked molecular clouds.

Reach and Rho \cite{RR99} suggest that the mass of compressed gas in
the shocked clump in 3C 391 is sufficiently high that self-gravity
is significant and that it could eventually form one or more stars.
Star formation in shocked supernova clumps
 cannot be directly verified because the clump collapse time is considerably
longer than the age of the supernova remnants.
By the time stars have formed, it is difficult to unambiguously discern
a supernova trigger.
One reason for interest in supernova triggered star formation has been
the evidence for extinct radioactivities in the early solar system
and this has stimulated computer simulations of the process
\cite{FB96,VC99}.
The observations of supernova remnants in molecular clouds can be
useful in providing realistic initial conditions for such simulations.

In cases where massive stars form in a dense cluster, the combined
effects of the photoionizing radiation and winds can effectively
clear molecular cloud material from the immediate vicinity of the stars.
However, observations of the Trapezium cluster in Orion have shown
the presence of small, dense gaseous regions, which O'Dell et al. \cite{O93}
have identified as proplyds (protoplanetary disks).
The eventual supernova explosion of one of the massive stars interacts with
the disk material; nearby disks are disrupted, but ones farther
out can survive \cite{C00}.
These interactions may have observable consequences in supernova remnants.

The study of supernova remnants in molecular clouds has attracted
renewed attention in the past few years and can be expected to 
blossom with the advent of new observatories and space missions,
such as the {\it ALMA} millimeter array, the {\it SOFIA} and
{\it SIRTF} infrared observatories, and the {\it GLAST} $\gamma$-ray
mission.
These observatories will make possible the detailed study of molecular
shock waves and $\gamma$-ray emission from relativistic particles.
The shock waves in molecular clouds provide an important probe
of  molecular cloud structure, which remains uncertain.
A combination of hydrodynamic studies with emission calculations
will be useful in elucidating this area.

\section*{acknowledgments}
This work was supported in part by NASA grants NAG5-8232 and NAG5-8088.

\end{document}